\documentclass[aps,pre,twocolumn,showkeys,showpacs,showdoi,amsfonts,amssymb,amsmath]{revtex4-1}
\usepackage{hyperref}
\usepackage{amsmath,amsfonts,amssymb}
\usepackage{graphicx}

\begin{document}

\title{On active information storage in input-driven systems}

\author{Oliver Obst} 
\affiliation{CSIRO Information and Communications Technology Centre, Epping, NSW 1710, Australia}
\author{Joschka Boedecker}  
\affiliation{Machine Learning Lab, University of Freiburg, 79110 Freiburg, Germany}
\altaffiliation[Part of the research conducted at ]{Graduate School of Engineering, Osaka University, Suita 565-0871, Osaka, Japan}
\author{Benedikt Schmidt}  
\affiliation{Artificial Intelligence Research Group, University of Koblenz-Landau, 56070 Koblenz, Germany}
\altaffiliation[Part of the research conducted at ]{Graduate School of Engineering, Osaka University, Suita 565-0871, Osaka, Japan}
\author{Minoru Asada}
\affiliation{Graduate School of Engineering, Osaka University, Suita 565-0871, Osaka, Japan}


\begin{abstract}
Information theory and the framework of information dynamics have been used to provide tools to characterise complex systems. In particular, we are interested in quantifying information storage, information modification and information transfer as characteristic elements of computation. Although these quantities are defined for autonomous dynamical systems, information dynamics can also help to get a ``wholistic'' understanding of input-driven systems such as neural networks. In this case, we do not distinguish between the system itself, and the effects the input has to the system. This may be desired in some cases, but it will change the questions we are able to answer, and is consequently an important consideration, for example, for biological systems which perform non-trivial computations and also retain a short-term memory of past inputs.
Many other real world systems like cortical networks are also heavily input-driven, and application of tools designed for autonomous dynamic systems may not necessarily lead to intuitively interpretable results. 

The aim of our work is to extend the measurements used in the information dynamics framework for input-driven systems. Using the proposed input-corrected information storage we hope to better quantify system behaviour, which will be important for heavily input-driven systems like artificial neural networks to abstract from specific benchmarks, or for brain networks, where intervention is difficult, individual components cannot be tested in isolation or with arbitrary input data.
\end{abstract}
%


\maketitle

\section{Introduction}
\label{sec:introduction}
In his 1990 paper~\cite{Lan90}, Langton addresses the question under what conditions physical systems support   the basic operations of information transmission, information storage, and information modification to support computation. In this investigation, cellular automata (CAs) are used as a formal abstraction of physical systems. Using a parameterisation of possible CA rules, a qualitative survey of the different dynamical regimes is presented, along with the observation that CAs exhibiting the most complex behavior are, in general, found near the phase transition between highly ordered and highly disordered dynamics. Information theory and the framework of information dynamics~\cite{LPZ08,LPZ10,Liz10} then provides the tools to quantify in complex systems the elements of computation using the basic operations information transmission, storage, and modification that have been mentioned above. These information-theoretic tools provide the means to understand, and to eventually engineer dynamical systems, a task for which a proper understanding of their computational properties is required. In contrast to static measurements of, e.g., entropy of a system at a given time, they focus on dynamical aspects of information processing. Understanding these dynamical aspects is critical and it has been suggested that ``the main challenge is understanding the dynamics of the propagation of information ... in networks, and how these networks process such information.''~\cite{Mit06}. 

Systems like CAs are autonomous dynamical systems, the evolution of their states at any given moment depends on a state-transition function and the current state. When instead dynamical systems are driven by some external input, the available tools may not be suitable to fully characterize them, and not lead to intuitively interpretable results: in this case, the information dynamics framework cited above, though useful to get a ``wholistic'' understanding of complex systems together with their input, will not necessarily provide useful information about the system in isolation. 
This is an important observation, as for example biological systems perform non-trivial computations and also retain a short-term memory of past inputs~\cite{GHS08}: using the information dynamics framework, there is no distinction between structure of the input into the system and that of the system itself. In some of our work, we have measured information transfer and active information storage in recurrent neural networks to show peak performance near the edge of chaos~\cite{BOL+11} for a number of inputs. 
Many real world systems like cortical networks are non-autonomous dynamical systems, and heavily input-driven. This requires new ways of investigating these systems, in particular if inputs are expected to change over time, and we are interested in their properties in face of change.

We are not the first identifying the need for new ways of analyzing non-autonomous dynamical systems: the work of 
Manjunath at al.~\cite{MTJ12} points out new developments in this area. Their focus are attractors, and how the concept translates to input-driven systems. Speaking about an example case, they ask ``Where does the perceived complexity of the state evolution come from? Is it due to the complex nature of the input driving source, or due to the complex autonomous dynamics of the individual maps [...], or both? Theory of autonomous systems, while profound and deep in many respects, is not suitable for answering such questions.''~\cite{MTJ12}.

In this paper, we attempt to provide the theory necessary to answer some of these questions for input-driven systems, starting from a basic concept that is used to quantify information storage in autonomous dynamical systems, the active information storage~\cite{LPZ12}. We extend this concept to the non-autonomous case, and illustrate that the computed quantities match the intuition of information storage using simple examples. 


\section{Active Information Storage}

 Active information storage, like information theory in general, has shown to be useful in general to analyze complex systems, and with it shares the advantage of being domain independent by using (Shannon) entropy as the fundamental quantity upon which it is based.  Before we give a definition of active information storage, we start with a few information-theoretical preliminaries. Entropy represents the uncertainty associated with any measurement $x$ of a random variable $X$, $H(X) = - \sum_x p(x) \log p(x)$, (where we use 2 as base for the logarithm, and \emph{bits} as unit for entropy). The conditional entropy of $X$ given $Y$ quantifies the amount of information needed to describe the outcome $x$ given that the value of $y$  is known: $H(X|Y) = - \sum_{x,y} p(x,y) \log p(x|y)$. The mutual information between $X$ and $Y$ measures the the average reduction in uncertainty about x that results from learning the value of y, or vice versa, and can be expressed via conditional entropies:
 \begin{align}
I(X;Y) & =H(X) - H(X|Y) = H(Y) - H(Y|X).  
 \end{align}
The conditional mutual information between X and Y given Z is the mutual information between X and Y when Z is known:
 \begin{align}
I(X;Y | Z) & =H(X|Z) - H(X|Y,Z).
 \end{align}

The concept of active information storage is derived~\cite{LPZ12} as the information in an agent, process or variableÕs past that can be used to predict its future. In contrast to excess entropy, which measures the total stored information that is used at some point in the future of the state process of an agent, the active information storage $A(X)$ expresses how much of the stored information is actually in use at the next time step  when the next process value is computed. $A(X)$ is expressed as the mutual information between the semi-infinite past of the process $X$ and its next state $X'$, with $X^{(k)}$ denoting the last $k$ states of that process:

\begin{align}
A(X) & = \lim_{k\to\infty} A^{(k)}(X) \\
A(X,k) & = I(X^{(k)}; X') \label{eq:kapprox}
\end{align}

Eq.~(\ref{eq:kapprox}) is also used to represent $k$-finite approximations of active information storage.

Active information storage is the \emph{average} amount of information in the past of a process that is in use to predict the next step, i.e., the expected value of the \emph{local} active information storage at each time step $n+1$. For a random variable $X$, the local active information storage for the value $x_{n+1}$ at time step $n+1$ is:
\begin{align}
a_{X}(n+1) & = \lim_{k\to\infty} a_{X}(n+1,k),\\
a_{X}(n+1,k) & = \log \frac{p(x^{(k)}_{n}, x_{n+1})}{p(x^{(k)}_{n}) p(x_{n+1})}
\end{align}

In a system of processes $\mathbf X$, the local active information storage for the value $x_{i,n+1}$ at time step $n+1$ of a process $i$ is defined as:
\begin{align}
a_{\mathbf X}(i,n+1) & = a_{X_i}(n+1), \\
a_{\mathbf X}(i,n+1,k) & = a_{X_i}(n+1,k).
\end{align}

With active information storage the average of its local values, we can write:
\begin{align}
A(X) & = \langle a_X(n+1) \rangle_{n}, \\
A(X,k) & = \langle a_X(n+1,k) \rangle_{n}. \label{eq:aisaveragek}
\end{align}

For sets of homogenous processes we can also average over all processes, i.e., 
\begin{align}
A({\mathbf X},k) & = \langle a_{\mathbf X}(i,n+1,k) \rangle_{i,n}. 
\end{align}

For details on the derivation and an in-depth discussion of active information storage and its properties we refer to~\cite{LPZ12}.

\section{Active Information Storage applied to input-driven systems}
\label{sec:aisinputdriven}

\begin{figure}[h]
\includegraphics[width=0.3\textwidth]{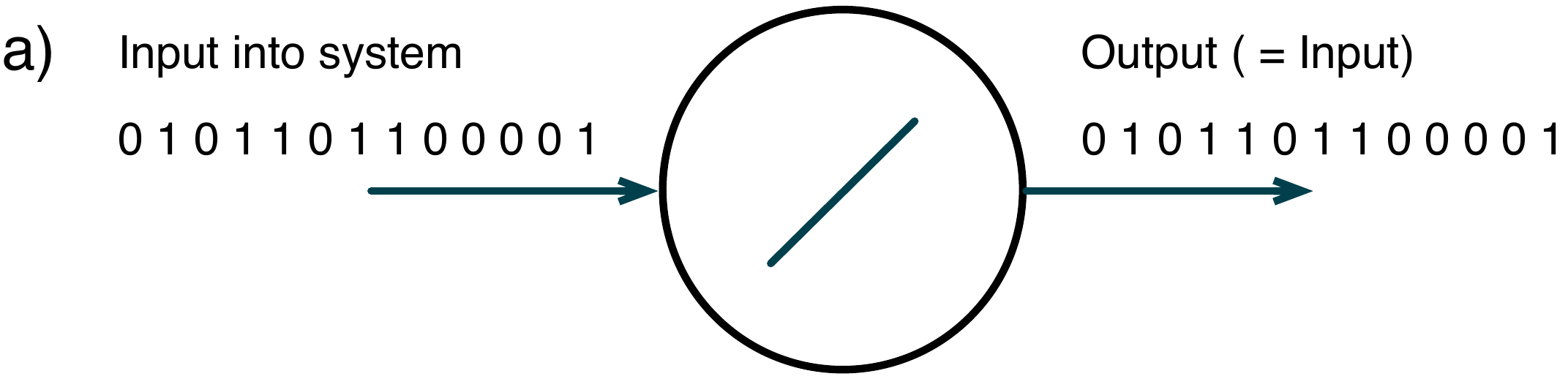}
\includegraphics[width=0.3\textwidth]{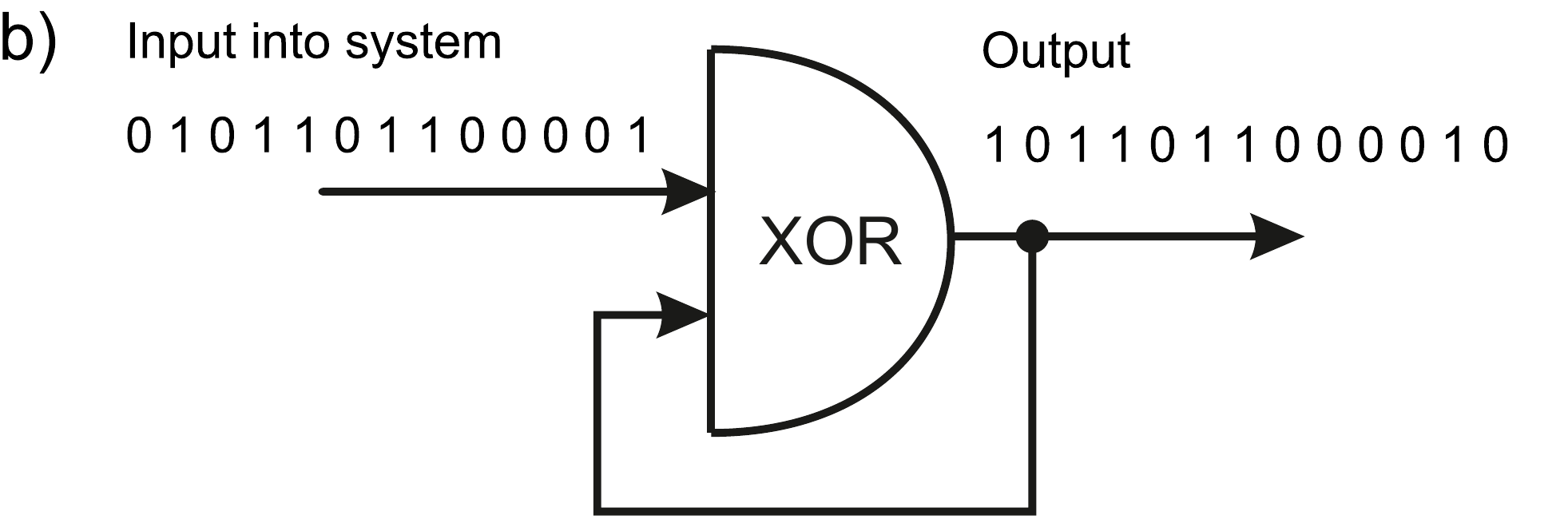}
\caption{Simple computational units of artificial neural networks may forward or store information. In a), inputs are just forwarded to the output.  b) implements a XOR-neuron that stores the last state to compute the output}
\label{fig:example1}
\end{figure}

To illustrate the effect of quantifying active information storage in an input-driven system, we look at two simple cases:   The first case (Fig.~\ref{fig:example1}a) is a simple forwarding unit, for which the output at step $n$ is the same as the input. In the second case (Fig.~\ref{fig:example1}b), the unit keeps its last output as an internal state. Its output is computed as logical xor between input and the internal state.

Intuitively,  in the first case we would expect zero active information storage for the unit, since no information is stored in the system. As we shall see, the computed active information storage will in fact depend on the structure of the input data. Similarly in the second case, we would expect one bit active information storage, since the units last state is required to compute its output. Again, we will see that the computed active information storage depends strongly on the structure of the input data. 

To demonstrate this effect we look at two specific kinds of input data, $u_1$ and $u_2$. For $u_1$ we draw values 0 and 1 independently from a Bernoulli distribution with $p = 0.5$. For $u_2$, we also draw binary random values, but impose a Markov condition so that with a probability of 0.7 the last value is repeated, and with a probability of 0.3, the value is changed from 0 to 1 or vice versa. 

Using these two time series to drive the forwarding unit, the probability of specific output values will be $p(x_n = 0) = p(x_n = 1) = 0.5$ in both cases, but the joint probabilities of two subsequent values will be different: For $u_1$, 
$p(x_n,x_{n+1}) = 0.25$, but for $u_2$, $p(x_n = x_{n+1}) = 0.7$, and $p(x_n \neq x_{n+1}) = 0.3$.

For a finite size approximation of active information storage with $k = 1$, the active information storage can be computed in both cases from the known (joint) probabilities (cf.~Eq.~\ref{eq:aisaveragek}), and evaluates as expected in the case of the i.i.d. input from $u_1$, to $A(X,1) = 0$, since $\log \frac {0.25} {0.5 \cdot 0.5} = 0$. It evaluates to, e.g., $A(X,1) \approx 0.1$, in the case of structured input from $u_2$, with $A(X,1) = 0.3 \log \frac {0.15} {0.5 \cdot 0.5} + 0.7 \log \frac {0.35} {0.5 \cdot 0.5}$. 

In the case of the xor unit, again using independent input data $u_1$, an output of 0 or 1 is equally likely independent of the current input: $p(x_n,x_{n+1}) = 0.25$. The computed active information storage for a history size of $k = 1$ will be zero. This is clearly counter-intuitive since the unit actually stores one bit of information that is required to compute its output. 

With increasing history sizes $k$, the computed values will eventually approximate the intuitively correct values of 0 and 1 respectively. Large history sizes, however, require large amounts of data to estimate the involved joint probabilities $p(x_n^{(k)}, x_{n+1})$. Oftentimes, the data required to produce reliable estimates are simply not available. With larger $k$ and larger data sets, estimation of $p(x_n^{(k)}, x_{n+1})$ becomes also more expensive. We aim to provide a solution using a new quantity that corrects the $k$-finite approximation of active information storage for input-driven systems.

\section{Active Information Storage for Input-Driven Systems}
\label{sec:aisids}

To correctly estimate active information storage for input-driven systems, we propose to condition out the input into the system. The local input-corrected active information storage at time step $n+1$ for a process $X$ with input $U$ thus becomes:

\begin{align}
a_{X}^{U}(n+1) & = \lim_{k\to\infty} a_{X}^{U} (n+1,k) \\
a_{X}^{U} (n+1,k) & = \log \frac {p(x^{(k)}_n, x_{n+1} | u_{n+1})} {p(x^{(k)}_n) p(x_{n+1} | u_{n+1})} \\ 
 & =  \log \frac {p(x_{n+1} | x^{(k)}_n, u_{n+1})} {p(x_{n+1} | u_{n+1})}
\end{align}

This measure can again be generalised to processes $X_i$ in a system $\mathbf X$:
\begin{align}
a_{\mathbf X}^U(i,n+1) & = \lim_{k\to\infty} a_{\mathbf X}^{U} (i, n+1,k) \\
a_{\mathbf X}^U(i,n+1,k) & = a_{X_i}^{U} (n+1,k) \\
 & = \log \frac {p(x_{i,n+1} | x^{(k)}_{i,n}, u_{n+1})} {p(x_{i,n+1} | u_{n+1})}.
\end{align}

We then have the {\bf input-corrected active information storage} $A_{\mathbf X}^U(i,k) = \langle a_{\mathbf X}^U(i,n,k) \rangle_n$. For homogenous processes we can again average over these, resulting in:
\begin{align}
 A_{\mathbf X}^U(k) = \langle a_{\mathbf X}^U(i,n,k) \rangle_{i,n}.
\end{align}

Applying the measure to our two example cases from above, we compute the respective conditional probabilities, again using a history size of $k=1$. In case of the forwarding unit, both local conditional probabilities $p(x_{n+1}|x_{n}^{(1)},u_{n+1})$ and $p(x_{n+1}|u_{n+1})$ evaluate to 1 for both the independent uniform input $u_1$ as well as for the structured input $u_2$, i.e., the input-corrected active information storage will be $\log 1 = 0$, independent of the input as we would expect.

In case of  the xor unit  conditioning on $u_{n+1}$ and $x_{n}^{(k)}$ leads to a probability of 1 for $p(x_{n+1}|x_{n}^{(1)},u_{n+1})$ while $p(x_{n+1}|u_{n+1}) = 0.5$ because of missing information about $x_{n}^{(1)}$. With these probabilities, the $A_X^U(1) = \log \frac{1}{0.5} = 1 $ for our second example, again independent of the input and exactly as we would expect.

\section{Relation of ICAIS to other measures}

ICAIS can be related to and expressed in terms of a number of other measures~\cite{McGill54,Bell03,WB10,LFW13}.

\subsection{Partial Information Decomposition}
\label{sec:pid}

\begin{figure}[htbp]
\centering
\includegraphics[width=0.4\textwidth]{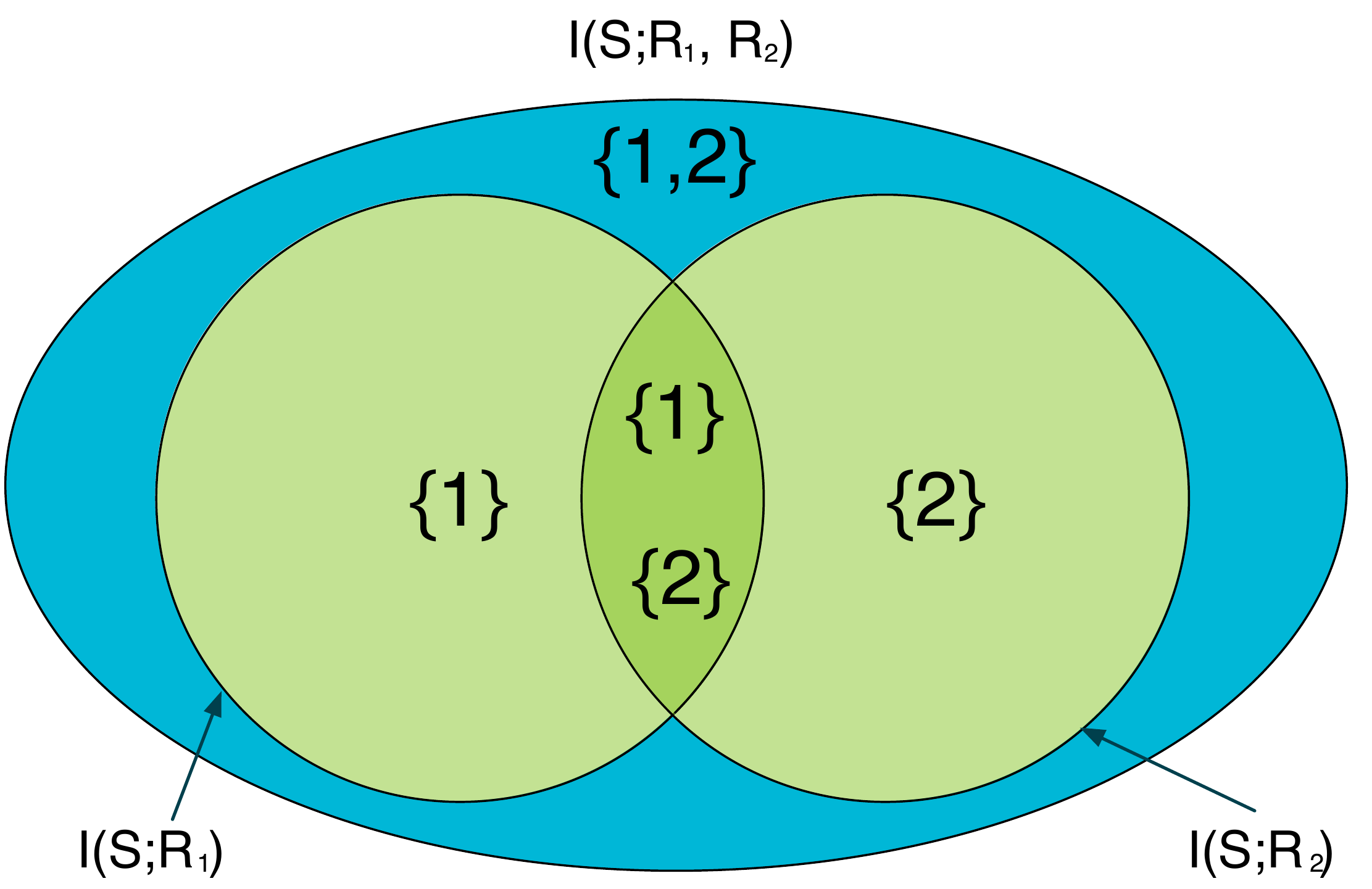}
\caption{Partial Information Decomposition for 3 variables.}
\label{fig:pid-3vars}
\end{figure}

Partial Information Decomposition (PID) is a recent framework~\cite{WB10} that decomposes information from several sources about a destination into information-theoretically atomic concepts of \emph{redundant}, \emph{unique} and \emph{synergistic} information. In the most simple case, for a system with three variables $S, R_1, R_2$, we want to know how much information provide $R_1$ and $R_2$ about $S$. It is possible to say how much $R_1$ and $R_2$ jointly contribute to the total information by using the mutual information $I(S; R_1,R_2)$. Decomposing this joint information, the amount of information that $R_1$ individually contributes (that is not found in $R_2$), or vice versa is the unique information. Information that is both in $R_1$ and in $R_2$ is called redundant information. The third concept, synergistic information, describes the situation when neither $R_1$ nor $R_2$ alone provide information about S but only jointly do so. Figure~\ref{fig:pid-3vars} visualizes the PID for the 3 variable case. The concept is not limited to 3 variables and can be applied to more complicated systems with any number of sources, $S = \{R_1, ..., R_n\}$. As nicely explained in \cite{LFW13}, it is defined in terms of an \emph{abstract}  method (in form of axioms that need to be satisfied), which needs an instantiation in form of a \emph{concrete} measure. 

\subsection{Interaction Information} 
\label{sec:interaction}

\begin{figure}[ht]
\centering
\includegraphics[width=0.4\textwidth]{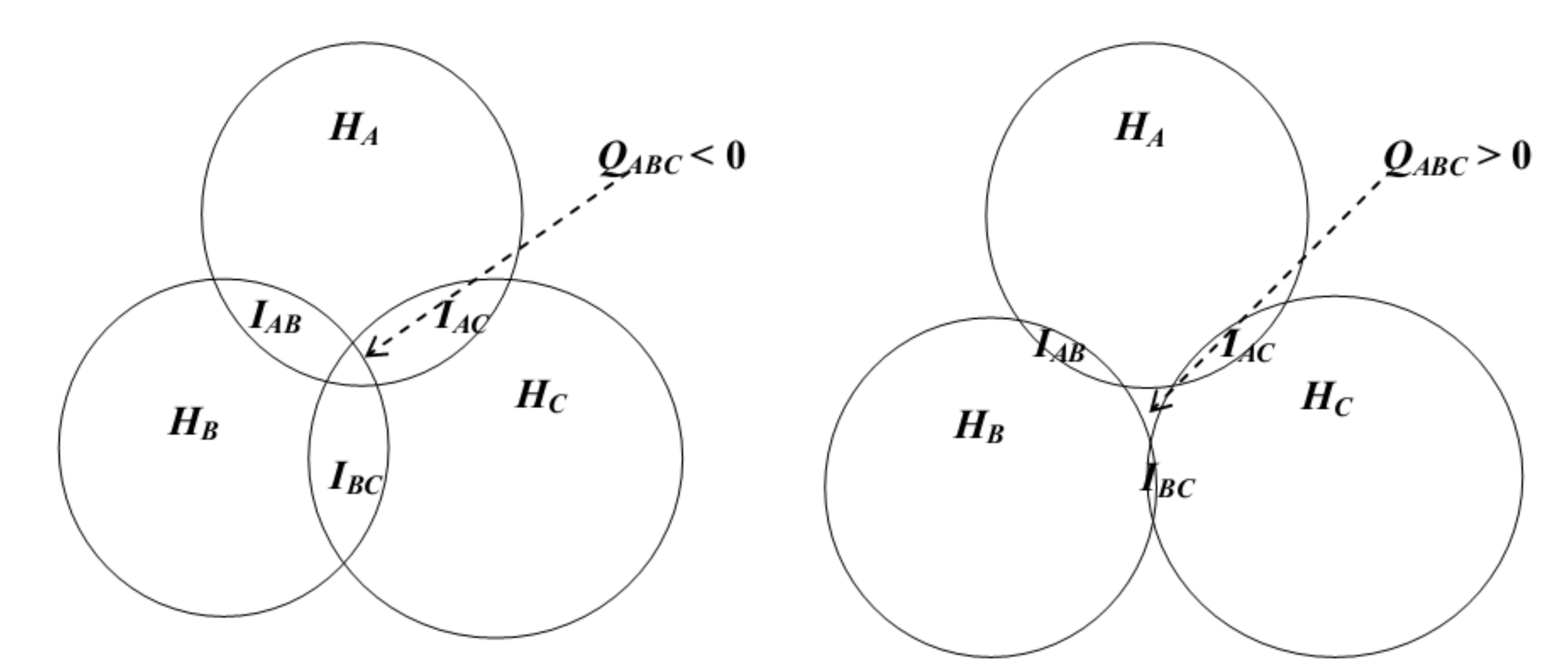}
\caption{Venn diagramm that visualizes the interaction information between three variables in case of redundancy (left) and synergy (right). \protect \cite{Leydesdorff09} }
\label{fig:interaction-info-example}
\end{figure}
Interaction information \cite{McGill54} or Co-Information \cite{Bell03} is a generalization of mutual information developed by McGill respectively Bell. It describes the information shared by k random variables, which can be positive or negative. 
The part of interest is the information shared between all three variables $I(X,Y,Z)$.  Here we want to show how this Idea is related to icAIS. Interaction Information for three variables is defined as follows:
\begin{align}
	I(X,Y,Z) 	& =\frac{I(X,Y|Z)}{I(X,Y)} \\
						& =\frac{I(X,Z|Y)}{I(X,Z)} \notag\\
						& =\frac{I(Y,Z|X)}{I(Y,Z)} \notag
\end{align}
where $I(X,Y|Z)$ and $I(X,Y)$ are defined as
\begin{align}
	I(X,Y|Z) 	& = log_2 \frac{p(X,Y|Z)}{p(X|Z)p(Y|Z)} \\
	I(X,Y)		& = log_2 \frac{p(X,Y)}{p(X)p(Y)}
\end{align}
As mentioned before interaction information can either be positive or negative for $k>=3$, what can be interpreted as synergy and redundancy \cite{Leydesdorff09}. If two sources contribute the same information to a destination redundancy occurs, this overlap is represented by a negative interaction information. In the opposite case of synergy and positive interaction information, two variables $U$ and $V$ contribute information that does not overlab. 
(see figure \ref{fig:interaction-info-example}) 
\newline
With icAIS we want to take redundancy and synergy explicitly into account. We can say we want to add the interaction that occurs between input and history to the AIS. We already see that $I(X,Y)$ equates to AIS, while equation \ref{eq:relation-i-and-icAIS} shows that $I(X,Y|Z)$ equates icAIS. 
\begin{align}
	I(X,Y|Z) 	& = \frac{p(X,Y|Z)}{p(X|Z)p(Y|Z)}  \text{substitute} X \notag \\
	 & = x_{n+1}, Y = x_n^{(k)}, Z = u_{n+1} \notag \\
	 & =\frac{p(x_{n+1},x_n^{(k)}|u_{n+1})}{p(x_{n+1}|u_{n+1})*p(x_n^{(k)}|u_{n+1})}  \notag \\
	 & = log \frac{p(x_{n+1},x_n^{(k)},u_{n+1})}{p(u_{n+1})} - log \frac{p(x_{n+1},u_{n+1})*p(x_n^{(k)},u_{n+1})}{p(u_{n+1})*p(u_{n+1})} \notag \\
	 & = log \frac{p(x_{n+1},x_n^{(k)},u_{n+1})*p(u_{n+1}}{p(x_{n+1},u_{n+1})*p(x_n^{(k)},u_{n+1})} \notag \\
	 & = log \frac{p(x_{n+1}|u_{n+1},x_n^{(k)})}{p(x_{n+1}|u_{n+1})} 
	 \label{eq:relation-i-and-icAIS}
\end{align}
Equation \ref{eq:relation-i-and-icAIS} proves that Interaction information can be written as $I = \frac{icAIS}{AIS}$, what can be transformed to $icAIS = AIS + I$ matching the asumption we made before. As it will be shown later in these thesis synergy and redundancy are the main issue applying AIS on a input driven system. 

\newpage

\bibliography{icaisrefs}

\end{document}